\documentclass[twocolumn]{revtex4}
\usepackage{amssymb,epsf}
\usepackage{latexsym}
\usepackage{xcolor}
\usepackage{float}
\begin{document}

\title{Gamow Temperature in Tsallis and Kaniadakis Statistics}
\author{Hooman Moradpour\footnote{h.moradpour@riaam.ac.ir}, Mohsen Javaherian, Ebrahim Namvar, Amir Hadi Ziaie}
\address{Research Institute for Astronomy and Astrophysics of Maragha (RIAAM), University of Maragheh, P.O. Box 55136-553, Maragheh, Iran}
\date{\today}
\begin{abstract}
Relying on the quantum tunnelling concept and Maxwell–Boltzmann–Gibbs statistics, Gamow shows that the star-burning process happens at temperatures comparable to a critical value, called the Gamow temperature ({\tt T}) and less than the prediction of the classical framework. In order to highlight the role of the equipartition theorem in the Gamow argument, a thermal length scale is defined, and then the effects of non-extensivity on the Gamow temperature have been investigated by focusing on the Tsallis and Kaniadakis statistics. The results attest that while the Gamow temperature decreases in the framework of Kaniadakis statistics, it can be bigger or smaller than {\tt T} when Tsallis statistics are employed.
\end{abstract}

\maketitle

\section{Introduction}

In the framework of Maxwell-Boltzmann-Gibbs statistics, also known as the Gibbs statistics, consider a gas with temperature
$T_g$, including particles with mass $m$ and mean velocity $v$. In this manner, using the equipartition theorem, one finds

\begin{eqnarray}\label{e1}
\frac{1}{2}mv^2=\frac{3}{2}K_BT_g.
\end{eqnarray}
\noindent Here, $K_B$ is the Boltzmann constant. For a pair of particles (called the $i$-th and $j$-th particles) with atomic numbers $Z_i$ and $Z_j$, respectively, located at a distance $r_0$ from each other, the Kinetic energy lets particles overcome the Coulomb barrier meaning that nuclear fusion begins, and consequently, a star is born if $\frac{3}{2}K_BT_g\geq U(r_0)=\frac{Z_iZ_je^2}{4\pi\epsilon_0r_0}$ leading to~\cite{1}
\begin{eqnarray}\label{e2}
T_g\geq\frac{Z_iZ_je^2}{6\pi\epsilon_0K_Br_0}\simeq 1\cdot1 \times
10^{10}\frac{Z_iZ_j}{r_0}\equiv T_0,
\end{eqnarray}
\noindent for the gas temperature. On the other hand, for the temperature of the gas, we also have~\cite{1}
\begin{eqnarray}\label{e3}
\mathcal{T}\approx4\times10^6\left(\frac{M}{M_\odot}\right)\left(\frac{R_\odot}{R}\right),
\end{eqnarray}

\noindent in which $M$ ($M_\odot$) and $R$ ($R_\odot$) denote the
mass and radius of the gas (Sun), respectively. As an example,
consider the Sun for which we have $\mathcal{T}\ll T_0$ meaning
that the Sun should not burn \cite{1}. Therefore, nuclear fusion
cannot be launched in gasses whose temperature (${\cal T}$) is lower than
$T_0$ (i.e., ${\cal T}<T_0$) \cite{1}.
Thanks to the scorching Sun, the above argument becomes questionable. Indeed, Gamow was the first one who was able to find a proper answer by proposing a mechanism: quantum tunneling~\cite{1}. Based on this theory, if the particles become close to each other as their de Broglie wavelength ($r_0\simeq\frac{\hbar}{p}\equiv\lambda$), then they overcome the Coulomb barrier. In this manner, the corresponding de Broglie wavelength of particles can be calculated as
\begin{eqnarray}\label{lambdaq}
\lambda=\frac{2\pi\epsilon_0\hbar^2}{mZ_iZ_je^2},
\end{eqnarray}
\noindent where $p=mv$ is considered, we replaced $r_0$ with $\lambda\equiv\frac{\hbar}{p}$ and then used
\begin{eqnarray}\label{lambda1}
\frac{p^2}{2m}=\frac{Z_iZ_je^2}{4\pi\epsilon_0\lambda}.
\end{eqnarray}
\noindent Now, using $\frac{3}{2}K_BT_g\geq U(r_0)$, one reaches~\cite{1}
\begin{eqnarray}\label{e4}
T_g\geq\frac{Z_iZ_je^2}{6\pi\epsilon_0K_B\lambda}\simeq 9\cdot6
\times 10^{6}Z_i^2Z_j^2\left(\frac{m}{\frac{1}{2}}\right)\equiv\texttt{T},
\end{eqnarray}

\noindent instead of Eq.~(\ref{e2}) meaning that nuclear fusion can be started in gases whose temperature are comparable with $\texttt{T}$ (the Gamow temperature) not $T_0$ \cite{1}. Moreover, using $\frac{3}{2}K_BT_g\geq U(r_0)$, one obtains
\begin{eqnarray}\label{lambdaqq}
\lambda\geq\frac{Z_iZ_je^2}{6\pi\epsilon_0K_BT_g}\equiv r_0^T.
\end{eqnarray}

\noindent Now, bearing the equal signs within Eqs.~(\ref{e1})
and~(\ref{lambda1}) in mind, we can finally deduce that the
minimum requirement for quantum tunneling in a gas with
temperature $T_g\geq\texttt{T}$ is $\lambda=r_0^T$. Hence, the
equipartition theorem has a vital role i.e., if it changes then
both Eq.~(\ref{e4}) and $r_0^T$ change.

Although extensivity is the backbone of the Gibbs statistics,
there are various arguments in favor of the non-extensivity,
especially in relativistic systems and those that involve
long-range interactions \cite{kanient,kani0,rev,sal,kani1,kani2}.
The Tsallis and Kaniadakis ($\kappa$) statistics are two of the
most famous and widely used generalized statistics frameworks
\cite{kanient,rev,sal,kani1,kani2} that propose generalized
versions of the equipartition theorem \cite{plaTsal,kani2,nunes}.
Motivated by various reasons such as the long-range nature of
gravity, and the probable relationship between the quantum aspects
of gravity and the non-extensivity \cite{nunes,epl,shab}, these
statistics have been employed to lead to notable outcomes in $i$)
describing dark energy \cite{nunes,epjc}, MOND theory \cite{mond},
$ii$) studying the Jeans instability \cite{jtsal,jk,jk1,morad},
and also $iii$) stellar sciences \cite{st1,st4,st2,st3,st5}.

Relying on the abovementioned achievements of Tsallis and Kaniadakis statistics, and the key role of the Gamow temperature in the stellar sciences, we are motivated to study the Gamow theory in these frameworks. Indeed, finding the Gamow temperature in Tsallis and Kaniadakis statistics is an important task that also helps one to obtain a better understanding of non-extensivity, gravity, and in fact, the non-extensive aspects of gravity. To achieve this goal, we focus on Tsallis and Kaniadakis statistics in the next section, and a summary will be presented at the end.

\section{Generalized statistics and the Gamow temperature}
\subsection{Tsallis framework}
The Tsallis entropy content of a statistical distribution with $W$
states while the $i$-th state happens with probability $P_i$ is
defined as \cite{sal}

\begin{eqnarray}\label{TSe}
S^T_q=\frac{1}{1-q}\sum_{i=1}^{W}(P_i^q-P_i),
\end{eqnarray}

\noindent where $q$ is a free parameter calculated by other parts
of physics or matching with experiments \cite{rev,sal}. The Gibbs
entropy is recovered at $q\rightarrow1$; in fact, each sample has
its own $q$ \cite{rev,sal}. For a three dimensional particle, the
ordinary thermal energy ($\frac{3}{2}K_BT$) is modified as
$\frac{3}{5-3q}K_BT$ meaning that Eq.~(\ref{e1}) changes as

\begin{eqnarray}\label{t1}
\frac{1}{2}m v^2=\frac{3}{5-3q}K_BT_g,
\end{eqnarray}

\noindent where $0\leq q <\frac{5}{3}$ \cite{nunes}. Now, simple
calculations lead to

\begin{eqnarray}\label{t2}
&&T_g\geq\frac{5-3q}{2}\texttt{T}\equiv\texttt{T}_q\Rightarrow 0<\texttt{T}_q\leq2\cdot5~\texttt{T},\nonumber\\
&&\lambda_q=\frac{2\pi\epsilon_0\hbar^2}{m Z_iZ_je^2}=\lambda,
\end{eqnarray}

\noindent in which the subscript $q$ is used to distinguish the
previous results with those of the Tsallis statistics. Moreover,
solving $\frac{3}{5-3q}K_BT_g=U(r_0)$, one reaches the Tsallis
thermal length scale
$r_{0q}^T(\equiv\frac{(5-3q)Z_iZ_je^2}{12\pi\epsilon_0K_BT_g}=\frac{5-3q}{2}r_0^T)$
that meets the condition $r_{0q}^T(q\rightarrow1)\rightarrow
r_0^T$. Therefore, quantum tunneling can happen at temperature
comparable to $\texttt{T}_q$, and in this manner, we should have
at least $r_{0q}^T=\lambda_q$.
\subsection{The $\kappa$ statistics}

In this framework, entropy is given by \cite{kani1}

\begin{eqnarray}\label{Ke}
&&S_\kappa=-\sum_{i=1}^{W}\frac{P_i^{1+\kappa}-P_i^{1-\kappa}}{2\kappa}=\\
&&\frac{1}{2}\bigg(\frac{\sum_{i=1}^{W}(P_i^{1-\kappa}-P_i)}{\kappa}+\frac{\sum_{i=1}^{W}(P_i^{1+\kappa}-P_i)}{-\kappa}\bigg),\nonumber
\end{eqnarray}

\noindent leading to \cite{epjc}

\begin{eqnarray}\label{Ke1}
&&S_\kappa=\frac{S_{1+\kappa}^T+S_{1-\kappa}^T}{2},
\end{eqnarray}

\noindent which clearly testifies that the Gibbs entropy is
achieved for $\kappa=0$ \cite{kanient}. Indeed, $\kappa$ is an
unknown free parameter estimated by observations that varies from
case to case \cite{kanient}. Moreover, the equipartition theorem
changes \cite{kani2,nunes}, and thus, Eq.~(\ref{e1}) takes the
form

\begin{eqnarray}\label{k1}
\frac{p^2}{2m}=\frac{3}{2}\gamma_\kappa K_BT_g,
\end{eqnarray}

\noindent in which

\begin{eqnarray}\label{k2}
\gamma_\kappa=
\frac{(1+\frac{\kappa}{2})\Gamma(\frac{1}{2\kappa}-\frac{3}{4})\Gamma(\frac{1}{2\kappa}+\frac{1}{4})}{2\kappa(1+\frac{3\kappa}{2})\Gamma(\frac{1}{2\kappa}+\frac{3}{4})\Gamma(\frac{1}{2\kappa}-\frac{1}{4})},
\end{eqnarray}

\noindent where $0\leq\kappa<\frac{2}{3}$ and $\Gamma(n)$ denotes
the Gamma function \cite{nunes}. Moreover, $\gamma_\kappa$
diverges for $\kappa=\frac{2}{3}$ and the ordinary equipartition
theorem ($\frac{3}{2}K_BT$), and thus Eq.~(\ref{e1}) are recovered
when $\kappa=0$ leading to $\gamma_\kappa=1$ \cite{nunes}.

Finally, it is a matter of calculation to find the Kaniadakis
counterpart of Eq.~(\ref{t2}) and the Kaniadakis length scale as

\begin{eqnarray}\label{k3}
&&T_g\geq\frac{\texttt{T}}{\gamma_\kappa}\equiv\texttt{T}_\kappa,\\
&&\lambda_\kappa=\frac{2\pi\epsilon_0\hbar^2}{m
Z_iZ_je^2}=\lambda,\nonumber
\end{eqnarray}

\noindent and

\begin{eqnarray}\label{k4}
r_{0\kappa}^T=\frac{r_0^T}{\gamma_\kappa},
\end{eqnarray}

\noindent respectively. Since $1\leq\gamma_\kappa$ \cite{nunes},
the conditions $\texttt{T}_\kappa\leq\texttt{T}$ and
$r_{0\kappa}^T\leq r_0^T$ are obtained as the allowed intervals
for $\texttt{T}_\kappa$ and $r_{0\kappa}^T$.

\section{Conclusion}

Reviewing the Gamow theory shows the role of equipartition theorem
in more clarification via defining a thermal length scale
($r_0^T$). It was deduced that the nuclear fusion would occur in a
gas whose temperature is comparable to the Gamow temperature
($\texttt{T}$) if the minimum requirement $\lambda=r_0^T$ is
satisfied. Moreover, equipped with the fact that generalized
statistics modifies the equipartition theorem and motivated by
their considerable achievements in various setups
\cite{nunes,epl,shab,epjc,mond,jtsal,jk,jk1,morad,st1,st4,st2,st3,st5},
we studied the Gamow temperature within Tsallis and Kaniadakis statistics. The results indicate that the Gamow temperature
($\texttt{T}$) decreases in the Kaniadakis statistics
($\texttt{T}_\kappa\leq\texttt{T}$), and in the Tsallis
statistics, it can be smaller or bigger than $\texttt{T}$ (i.e.,
$0<\texttt{T}_q\leq2\cdot5~\texttt{T}$), depending on the value of
$q$. The same result applies to the corresponding thermal length
scales.
\par
Correspondingly, it may be claimed that stars whose temperature ${\cal T}$ differs from ${\tt T}$ are signals of the non-extensive features of stellar sciences, meaning that if stars obey Tsallis (Kaniadakis) statistics, then by using ${\cal T} = {\tt T}_q ({\cal T} = {\tt T}_\kappa)$, one can find the value of $q\,\,(\kappa)$ corresponding to each star. Hence, the upper and lower bounds on the $q\,\,(\kappa)$ parameter for nuclear fusion process occurring can be found in the coldest and hottest stars. Finally, we should note that further theoretical studies and also fitting with observations are needed to determine the final probable generalized statistics governing the stellar sciences.
\par
{\bf Data Availability Statement:} This manuscript has no associated data or the data will not be deposited.

\end{document}